\pdfoutput=1

\documentclass[11pt]{article}

\usepackage[final]{coling}

\usepackage{times}
\usepackage{latexsym}

\usepackage[T1]{fontenc}

\usepackage[utf8]{inputenc}

\usepackage{microtype}

\usepackage{inconsolata}

\usepackage{graphicx}
\usepackage{tikz}
\usepackage{booktabs}
\usepackage{rotating}
\usepackage[normalem]{ulem}
\usepackage{multicol}
\usepackage{multirow}
\usepackage{adjustbox}
\usepackage{amsmath,amssymb,amsfonts}
\usepackage{algorithmic}
\usepackage{textcomp}
\usepackage{pgfplots}
\usepackage{pgfplotstable}
\usepackage{hyperref}
\usepackage{inconsolata}
\usepackage{xcolor}
\pgfkeys{/pgf/number format/.cd,1000 sep={}}
\pgfplotsset{compat=1.17}
\usetikzlibrary{patterns}
\usepackage[normalem]{ulem}
\usepackage{natbib}
\usepackage{pifont}
\usepackage{arydshln} 
\usepackage{pdflscape}
\usepackage{fancybox}
\usetikzlibrary{shapes.geometric, arrows}
\usepackage[most]{tcolorbox}
\usepackage{inconsolata}
\usepackage{subfig}

\title{No Size Fits All: The Perils and Pitfalls of Leveraging LLMs Vary with Company Size}

\author{Ashok Urlana\textsuperscript{1} \enspace \enspace Charaka Vinayak Kumar\textsuperscript{1} \enspace \enspace  Bala Mallikarjunarao Garlapati\textsuperscript{1} \\ \enspace \enspace \textbf{Ajeet Kumar Singh\textsuperscript{1}} \enspace \enspace \textbf{Rahul Mishra\textsuperscript{2}}\\
TCS Research, Hyderabad, India\textsuperscript{1} \enspace \enspace \enspace \enspace \enspace \enspace
IIIT Hyderabad\textsuperscript{2}\\
{\tt ashok.urlana@tcs.com}, {\tt charaka.v@tcs.com}, {\tt balamallikarjuna.g@tcs.com} \\{\tt ajeetk.singh1@tcs.com}, {\tt rahul.mishra@iiit.ac.in}
}

\begin{document}
\maketitle
\begin{abstract}
Large language models (LLMs) are playing a pivotal role in deploying strategic use cases across a range of organizations, from large pan-continental companies to emerging startups. The issues and challenges involved in the successful utilization of LLMs can vary significantly depending on the size of the organization. It is important to study and discuss these pertinent issues of LLM adaptation with a focus on the scale of the industrial concerns and brainstorm possible solutions and prospective directions. Such a study has not been prominently featured in the current research literature. In this study, we adopt a threefold strategy: first, we conduct a case study with industry practitioners to formulate the key research questions; second, we examine existing industrial publications to address these questions; and finally, we provide a practical guide for industries to utilize LLMs more efficiently. We release the GitHub\footnote{\url{https://github.com/vinayakcse/IndustrialLLMsPapers}} repository with the most recent papers in the field.
\end{abstract}

\section{Introduction}
Large language models (LLMs) have recently garnered significant attention due to their exceptional performance in various predictive and generative tasks \cite{hadi2023survey, kar2023unravelling}. Extensive research has been conducted to harness LLMs across diverse domains and tasks \cite{raiaan2024review}, including medicine \cite{thirunavukarasu2023large}, finance \cite{paper_051}, and reasoning tasks \cite{huang-chang-2023-towards, qiao-etal-2023-reasoning}. Despite their unprecedented adaptation to numerous industrial applications, there is a notable lack of studies examining the potential challenges and risks associated with LLMs, which can vary depending on the size of the organization. Such studies would not only be valuable for industries seeking informed adaptation but also help shape research focus to address the key challenges and obstacles faced in real-world scenarios.

The challenges and bottlenecks faced by organizations of different sizes are not uniform. Factors such as funding availability, workforce size, skill and training deficits, ethical and regional considerations, and access to adequate hardware can all influence how these challenges manifest. Previous research has largely addressed general challenges \cite{raiaan2024review} with LLMs, such as multilingual support, domain adaptation, and compute requirements. However, there is a lack of studies specifically focusing on the industrial perspective and the unique challenges of implementing LLMs in this context.

To this end, we conduct a study with a threefold strategy, firstly, we conduct a rigorous case study of real-world practitioners from the IT industry, who are trying to work on AI adaptation and formulate three guiding research questions. \noindent \textbf{\textit{RQ1.}} How have industries adopted LLMs so far, and what challenges do they face? 
 \textbf{\textit{RQ2.}} What are the barriers hindering the full utilization of LLMs in industrial applications, and how can these barriers be addressed? 
 \textbf{\textit{RQ3.}} How can various industries advance to maximize the utility of LLMs in practical applications? Subsequently, with an aim to address guiding research questions, we perform a thorough scoping survey of existing research publications from industrial entities of all sizes. Finally, we discuss our takeaways and insights and present a practical pilot scenario-based guide for industries to adapt to LLMs in a more informed manner.

The key contributions of this work can be summarised as: this study identifies various categories of challenges associated with LLMs for industrial adoption and proposes potential solutions. These challenges broadly relate to data confidentiality, reliability of LLM responses, infrastructure bottlenecks across industries, domain-specific adoption, synthetic data generation, and ethical concerns. Additionally, we offer a practical guide tailored for small, medium, and large industries to maximize the utilization of LLMs.   


\section{Related Work}

In the literature, numerous studies focus on practical and ethical challenges associated with LLMs across diverse application domains includes education \cite{yan2024practical}, finance \cite{li2023large}, healthcare \cite{zhou2023research} and security \cite{shao2024empirical}. Additionally, several studies address the task-specific challenges for LLMs' adoption in areas such as spoken dialog systems \cite{inoue2023challenges}, mathematical reasoning \cite{ahn2024large}, mining software repositories \cite{abedu2024llm}. Moreover, studies explore the challenges based on LLMs capabilities with explanations generation \cite{kunz-kuhlmann-2024-properties}, data augmentation \cite{ding2024data}, support for multilingual context \cite{shen2024language} and compliance with ethical challenges \cite{jiao2024navigating}.

Close to our work, \citet{gallagher2024assessing} addresses a few concerns on the adoption of LLMs for specific high-stake applications, particularly intelligence reporting workflows. In contrast to existing studies, our work specifically concentrates on the utilization of LLMs for industrial applications. Moreover, this study provides a comprehensive overview of several roadblocks to LLMs adoption for industrial use cases and corresponding potential solutions. Additionally, our study offers a suggestive guide to maximize the utilization of LLMs for various industries.


\section{Methodology}
This section aims to explore how industries have adopted LLMs and the challenges they face (\textbf{RQ1}).
\subsection{Industrial Case Study on LLMs}
\label{sec:industry_survey}
We conduct an industrial case study to understand, how the LLMs are shaping industry practices, identify the underlying challenges and benefits. Through a meticulous process of expert consultation and iterative refinement, the questionnaire was designed to capture insightful data and serve as a tool for understanding the evolving role of LLMs in the industry. This case study covers a multitude of aspects related to LLM usage for specific application domains, corresponding risks, trust attributes, and challenges. In crafting a succinct questionnaire, our objective was to gauge the adoption and impact of LLMs in various industries. These questions can be found in Appendix~\ref{sec:case_study} Table~\ref{tab:survey_q2}. We receive 26 responses in total from real-world practitioners of the IT industry. We did a case study on 26 companies which are leveraging LLMs for their use-cases. This exercise is non-trivial as most companies have not made their LLM-related use cases public.
\setlength{\fboxrule}{0pt} 
\begin{figure*}%
    \centering
    \fbox{%
        \begin{minipage}{\textwidth}
            \centering
            \subfloat[\centering Data modalities considered.]{{\includegraphics[width=3.6cm]{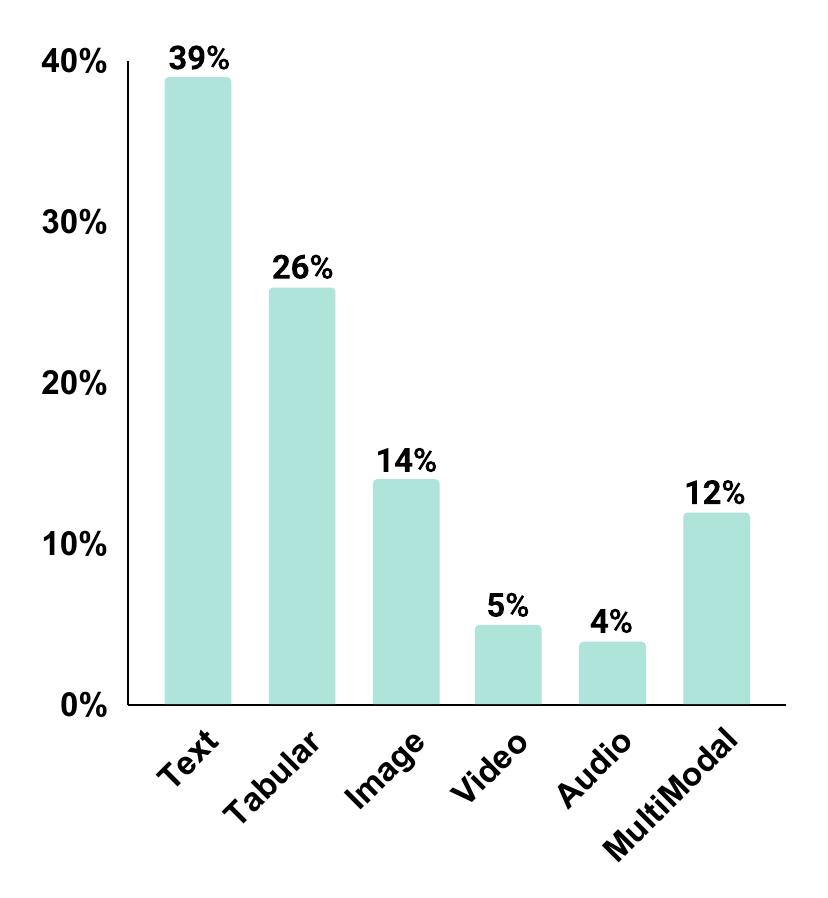}}}%
            \qquad
            \subfloat[\centering Foreseen risks associated with LLMs.]{{\includegraphics[width=3.6cm]{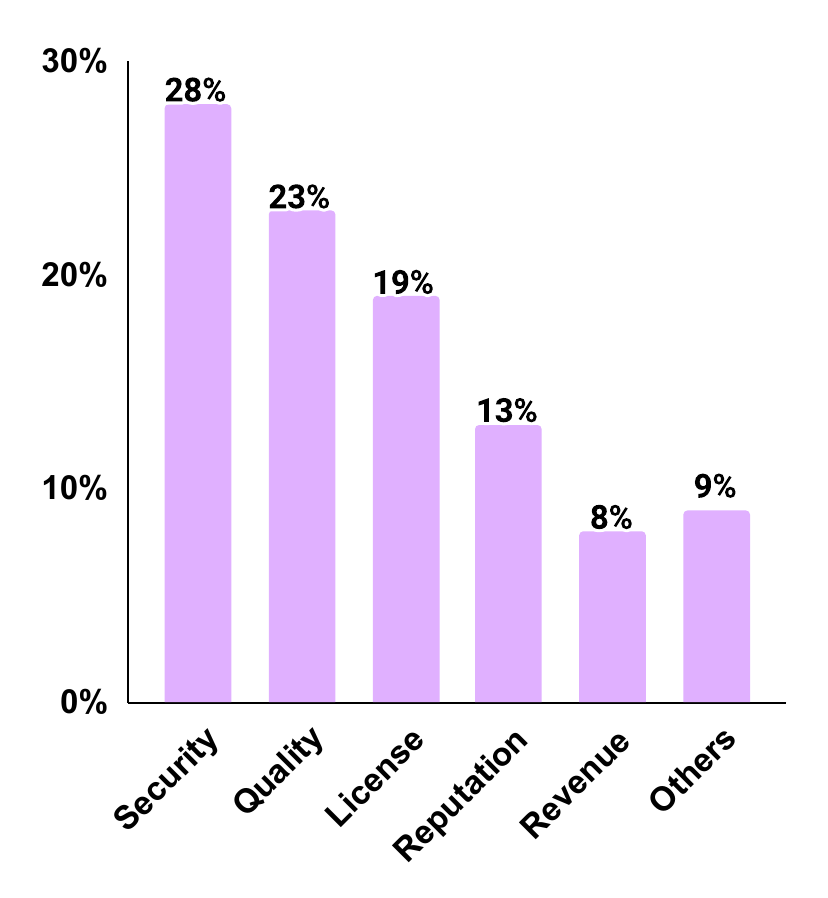}}}%
            \subfloat[\centering Opted trust attributes.]{{\includegraphics[width=3.6cm]{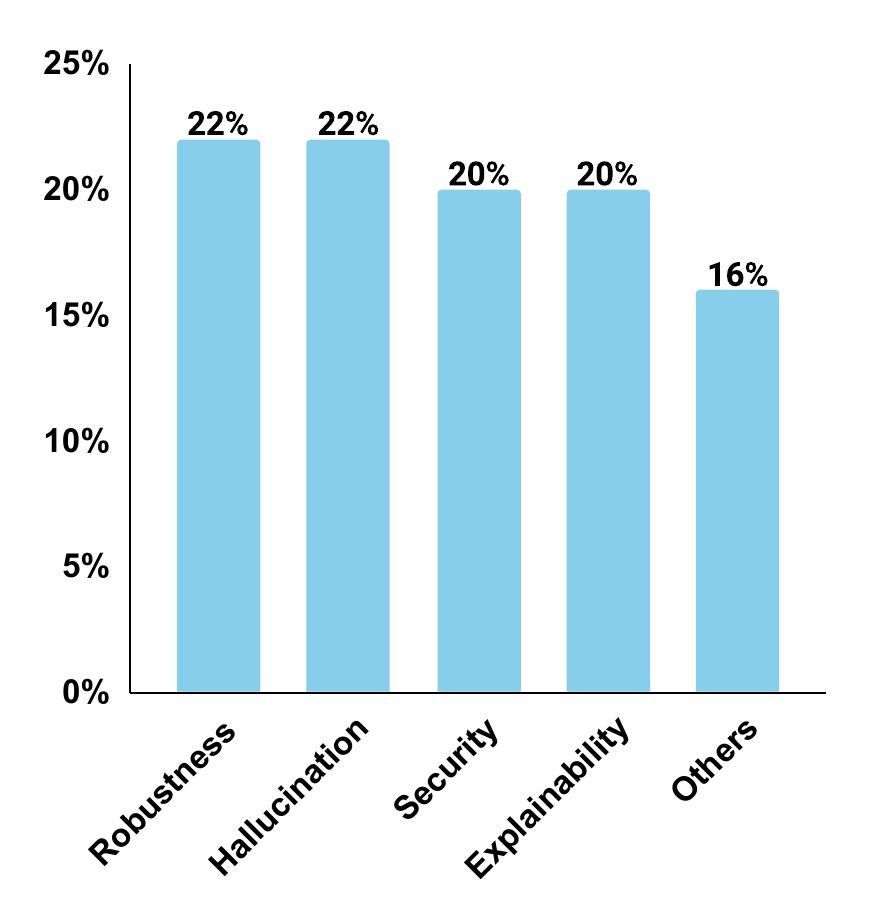}}}%
        \end{minipage}
    }
    \caption{Industrial case study statistical overview of various aspects}%
    \vspace{-5mm}
    \label{fig:case_study}
\end{figure*}
\subsection{Quantitative Analysis}
Based on the responses obtained from the industrial case study, we make the following observations.\\
\colorbox{gray!20}{Participants of the case study.} We shared the questionnaire with the IT professionals, who are either working on LLMs or have developed some solutions. The participants are industry professionals and practitioners with expertise ranging from beginner to expert level.\\
\colorbox{gray!20}{Widely adapted applications by leveraging LLMs.} Even though LLMs are being utilized for various applications, we observe that the majority of these industrial applications are related to financial, retail, security, and healthcare domains.\\
\colorbox{gray!20}{Modality of the datasets.} More than 60\% of the industry practitioners prefer to use either textual or tabular data as shown in Figure~\ref{fig:case_study}.a. \\
\colorbox{gray!20}{Widely used LLMs.} Our case study indicates more than 50\% of the applications utilize the GPT-3.5 and GPT-4 models. Recently, researchers have been assessing the capabilities of LLaMA-2 \cite{touvron2023llama} and Mistral \cite{jiang2023mistral}. \\
\colorbox{gray!20}{Prompting strategy.} We observe that zero-shot and in-context learning prompting strategies are widely adapted compared to fine-tuning.\\
\colorbox{gray!20}{Risks associated with LLMs.} Based on our case study, LLMs pose risks associated with security and safety, quality of service, and license-related challenges as depicted in Figure~\ref{fig:case_study}.b. \\
\colorbox{gray!20}{Trust attributes to be considered.} We observe that robustness, security, and hallucination are the major challenges that need to be considered to utilize the LLMs as shown in Figure~\ref{fig:case_study}.c.

Moreover, to gain a better understanding of the barriers to leverage the LLMs for industrial use cases, we also survey 68 research papers specifically from the industry. In this study, we compile several prominent challenges and present potential solutions to address them. The selection criteria for the papers can be found in the Appendix~\ref{sec:papers_selection}.

\section{Challenges and Potential Solutions}
 \label{sec: section_4}
In this section, we explore several barriers to leveraging LLMs for industrial applications and discuss potential solutions \textbf{(RQ2)}. 
\subsection{Data Confidentiality}
\subsubsection{Pre-training data issues}
\textbf{Potential privacy risks.} To deploy large language models (LLMs) on cloud platforms, robust data privacy protocols are required to handle extensive sensitive datasets while pre-training. Key challenges include mitigating data breaches and preventing unauthorized extraction of sensitive information. Despite the adoption of LLMs in applications like disaster response management \cite{paper_116}, public health intervention \cite{paper_057}, and assisting Augmentative and Alternative Communication (AAC) users \cite{paper_059}, there is noticeable lack of focus on privacy and security aspects. Moreover, it is imperative that potential risks associated with deploying LLMs in high-stakes scenarios are addressed. \\
\textbf{Regulations.} GDPR in Europe and CCPA in California introduce stringent guidelines for deploying LLMs by enforcing strict data handling and intellectual property rules to ensure transparency and fairness. 
As highlighted by \citet{reg_med}, adhering to these laws in sensitive domains like healthcare is crucial to avoid harm and protect privacy.\\
\colorbox{gray!20}{\textbf{Potential solution.}} Developing a comprehensive framework that aids in LLM compliance is essential for responsible use and interaction with users.
\subsubsection{Usage of APIs}
To access the closed-source LLMs, passing the commercial data through third-party APIs raises potential privacy concerns \cite{paper_004}.\\
\noindent \colorbox{gray!20}{\textbf{Potential solutions.}} 1. Robust security and privacy techniques like federated learning are essential to safeguard user data while maintaining the functionality of LLMs, 2. A strategic way of crafting prompts is essential to avoid Personally Identifiable Information (PII) leakage \cite{paper_056}. 
\subsection{Reliability of LLMs' Responses}
\textbf{Control the level of AI proactivity.} LLMs should minimize social awkwardness, enhance expressiveness, and adapt to different scenarios \cite{paper_047, urlana2023controllable}. The open-ended generation of LLMs makes it challenging to customize dialog systems for public health intervention applications \cite{paper_057}.\\
\textbf{Outdated knowledge.} The open-endedness of LLMs often leads to hallucinations due to a lack of an updated knowledge base \cite{faizullah2024limgen}. Additionally, the training data might contain errors and become outdated over time. \\
\colorbox{gray!20}{\textbf{Potential solutions.}} Techniques such as Retrieval-Augmented Generation are effective in reducing hallucinations. However, such systems struggle with complex questions that require additional information often generating out-of-context content. Moreover, techniques such as diverse beam search \cite{vijayakumar2016diverse}, confident decoding \cite{tian2019sticking} are promising in mitigating hallucinations. Additionally, model editing techniques \cite{hoelscher-obermaier-etal-2023-detecting} can address the unintended associations, enhancing the practical usage of LLMs.

\subsection{Infrastructure Accessibility}
\textbf{Carbon emissions.} Infrastructure is crucial for deploying LLMs, influencing factors like processing speed, latency, cost, and training needs. High-performance hardware is necessary to boost speed and reduce latency, enhancing user experience but it requires careful budgeting due to associated high costs. Achieving an optimal balance between cost and performance is crucial for the efficient training and scalability of LLM applications. \\
\colorbox{gray!20}{\textbf{Potential Solution.}} Implementing robust small language models lead to reduced carbon emissions. \\
\textbf{Compute requirements.} Despite the state-of-the-art performance of the large language models, utilizing them for small-scale industries is not feasible due to high compute requirements. \\
\textbf{API costs.} While LLMs like GPT-3.5 and GPT-4 \cite{achiam2023gpt} demonstrate superior performance over open-source models, their high cost of API access is prohibitively expensive to perform comprehensive studies \cite{paper_004}.\\
\colorbox{gray!20}{\textbf{Potential Solution.}} Balancing the trade-off between performance and cost is necessary for the practical usage of LLMs \cite{paper_004}.  \\
\textbf{High inference latency.} APIs can be slow when demand is high. For instance, tasks like business meeting summarization can take GPT-4 around 40 seconds to generate a single response \cite{paper_004}. Additionally, longer prompts increase computational demand \cite{jiang-etal-2023-llmlingua}. \\
\colorbox{gray!20}{\textbf{Potential Solution.}} Open-source models like LLaMA-2 \cite{touvron2023llama} are more favorable for industrial deployment. Further studies on efficient model optimization techniques such as quantization, pruning, and distillation are required \cite{paper_004}. Moreover, closed-source models that can utilize prompt compression techniques such as LLMLingua \cite{jiang-etal-2023-llmlingua}.

\subsection{Domain Adaption}
\textbf{Lack of domain-specific datasets.} The ability of LLMs in the finance and medical domains is lacking due to insufficient domain-specific training data in the foundation models \cite{paper_107, li-etal-2023-chatgpt}. Consequently, the current versions of GPT-4 and ChatGPT do not meet the industrial requirements to build financial analyst agents \cite{li-etal-2023-chatgpt}. While LLMs can generate relevant reasoning, they fall short of the desired standard, indicating significant room for improvement. 
 \\
\textbf{Diversity.} LLMs fail to mitigate social bias due to a lack of diverse demographic data \cite{paper_010}. Foundation models must equally consider factors like ethnicity, nationality, gender, and religion, as most currently reflect western perspectives.\\
\textbf{In-context learning (ICL).}  The scope of in-context learning is limited by its pretraining data \cite{han-etal-2023-understanding}. It is unlikely that any model will perform well when using ICL with data significantly different from its pretraining data.\\
\colorbox{gray!20}{\textbf{Potential Solution.}}  1. Pre-training data should consist of various domain mixtures; however, finding the right mixture is still an open challenge. 2. LLMs should be carefully tested to ensure they treat marginalized individuals and communities equally \cite{paper_027}. 3. Continuous pre-training can help overcome the drawbacks of the in-context learning strategy.    

\subsection{Data Creation Using LLMs}
Few works attempt to generate synthetic datasets by utilizing LLMs. However, three major concerns exist with using LLMs for synthetic data creation/annotation; 1). \textbf{Lack of diversity.} Synthetic datasets may lack diversity due to the limited knowledge base \cite{paper_019} of LLMs, 2). \textbf{Quality and compute.} The quality of the annotated data might improve with the size of the LLM used for the annotation \cite{paper_026}. However, leveraging large LLMs requires higher computational resources, 3). \textbf{In-context learning (ICL) challenges.} ICL is a widely adopted approach for textual task data annotation tasks \cite{li-etal-2023-chatgpt}. However, the main challenge lies in responsibly incorporating the model's output is to deliver value to users without misleading them or inadvertently amplifying malicious behavior \cite{paper_099}.\\
\colorbox{gray!20}{\textbf{Potential solution.}} Currently, tools like FABRICATOR \cite{paper_064}, support tasks like classification, sentence similarity and QA for data labeling and other tasks should be explored. 

\subsection{Sub-standard Performance of LLMs}
\textbf{Code generation.} LLMs' coding ability is limited to generate general-purpose coding tasks. However, the generation of high-quality code for complex network management tasks remains challenging \cite{paper_036}. Moreover, LLMs have limited capabilities in repository-level coding tasks except in C and Python languages \cite{paper_035} and fail to complete code with potential bugs \cite{paper_100}. Most of the code-LLMs struggle with code completion tasks, with undefined names and unused variables \cite{paper_053} being the most prominent static error cases.\\
\textbf{Conversational applications.} LLMs face challenges in providing emotional support and maintaining long-term memory, impacting their effectiveness in conversational applications \cite{paper_057}. Future research on a longitudinal deployment of LLM-driven chatbots for public health interventions would help understand how users’ engagement changes over time. \\
\textbf{Multilingual and Multi-Modal:} Most of the LLMs are being limited to English, there is significant room for creating robust multilingual models. Only a few studies have focused on utilizing LLMs for such multi-modal industrial applications \cite{paper_049, paper_052}. More efforts are needed to integrate LLMs with voice assistants and Robotics \cite{paper_078}. 

\subsection{Explainability and Interpretability}
The robust performance of LLMs across various tasks underscores the importance of explainability and interpretability to foster trust in their predictions. However, several challenges impede the development of explainable models.\\
\textbf{Black Box Nature:} Many popular LLMs, such as ChatGPT and Gemini \cite{team2023gemini}, are accessible only through APIs, limiting users' understanding of their internal workings.\\
\textbf{Scale and Complexity of Models:} The large-scale training on vast data leads to complex models, making it hard to identify which parameters influence specific decisions \cite{GPT3pt5}.\\
\textbf{Performance Trade-Off:} Balancing model performance with the ability to provide meaningful explanations is a significant challenge; many models struggle to maintain this equilibrium.\\
\textbf{Language Ambiguity:} The inherent ambiguity of language complicates the generation of clear explanations, as words and sentences can have multiple meanings depending on context \cite{wang2023deciphering}.\\
\colorbox{gray!20}{\textbf{Potential Solutions.}}
\textbf{Model Simplification:} Developing simpler models can enhance interpretability, provides a clear understanding of the decision-making processes of LLMs \cite{che2016interpretable}.\\
\textbf{Training Data Transparency:} Sharing details about training datasets and their sources can illuminate knowledge gaps and potential biases in the models \cite{bender2018data}.\\
\textbf{Interactive Exploration Tools:} Creating interactive platforms that allow users to manipulate inputs, visualize attention patterns, and observe changes in outputs can provide valuable insights into model behavior \cite{olah2018building}. 
\begin{figure}[t]
    \centering
    \includegraphics[width=0.33\textwidth]{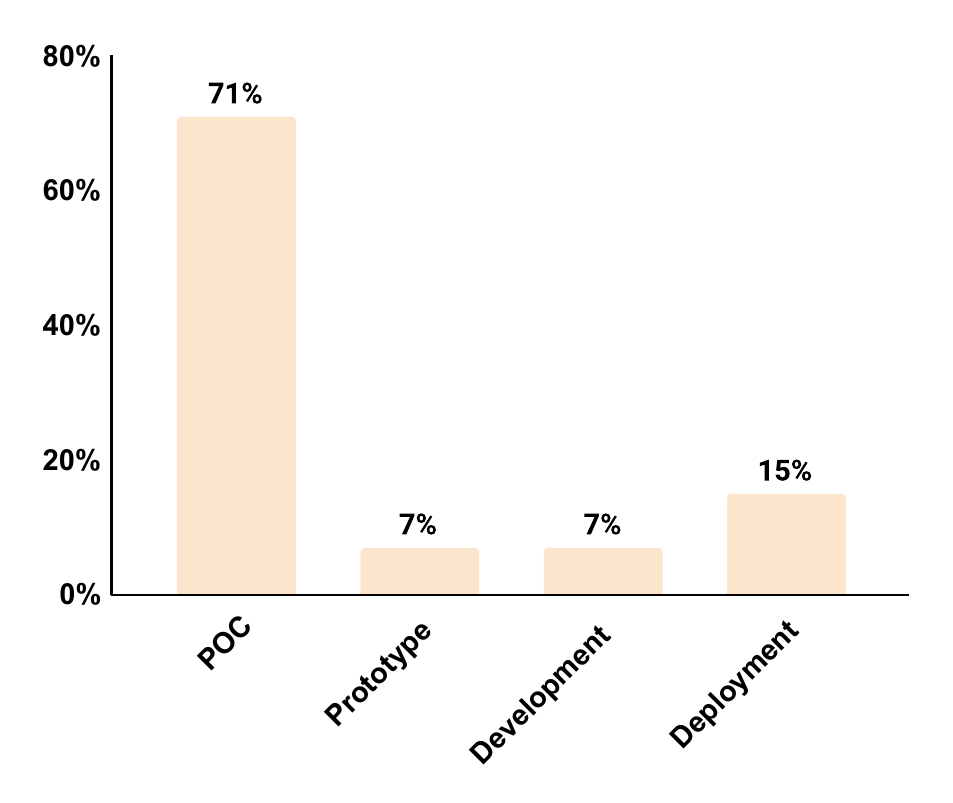}
    \vspace{-5mm}
    \caption{Current state of the industrial applications utilizing the LLMs; POC stands for proof of concept.}
    \vspace{-5mm}
    \label{fig:life_cycle_fig}
\end{figure}
\subsection{Evaluation of LLMs}
In sectors like legal, finance, and healthcare, blending LLMs with human feedback is crucial to lowering false positives, underscoring the importance of human oversight in safety-critical applications \cite{paper_090}. Moreover, our analysis (see Appendix~\ref{sec:check_list_survey}) reveals that less than 15\% of studies conduct human evaluations to assess LLM outputs, indicating a need for more rigorous validation methods. Evaluating long-form question answering is challenging for LLMs \cite{paper_002}, as additional contextual information may not always be available in practical QA scenarios. Current metrics like BLEU \cite{papineni-etal-2002-bleu} and ROUGE \cite{lin-2004-rouge} primarily evaluate the similarity, but are insufficient for assessing the reasonableness of LLM responses.
\begin{table*}[htb]
\centering\footnotesize
\setlength{\tabcolsep}{0.6ex}
\resizebox{\textwidth}{!}{%
\begin{tabular}{@{}llll@{}}
\toprule
 &
  \textbf{Small Scale Industries} & \textbf{Medium Scale Industries} & \textbf{Large Scale Industries} \\ \midrule
\textbf{\begin{tabular}[c]{@{}l@{}}Mode of LLMs \\usage \end{tabular}} & \begin{tabular}[c]{@{}l@{}}API Integration, Pre-trained Models, \\ Low-code or No code platforms, \\ Zero-shot, Few-shot\end{tabular} & \begin{tabular}[c]{@{}l@{}}Domain-specific Fine-tuning, \\ Chain of thought\end{tabular} &
  \begin{tabular}[c]{@{}l@{}}In-house deployment, Continuous pre-training, \\ Collaborative tools and frameworks, \\ Pre-training from scratch, Re-pretraining\end{tabular} \\ \midrule
\textbf{Challenges} &
  \begin{tabular}[c]{@{}l@{}}Cost, Technical expertise, Data\\ privacy, Performance\end{tabular} &
  \begin{tabular}[c]{@{}l@{}}Scalability, \\ Domain Adoption\end{tabular} &
  \begin{tabular}[c]{@{}l@{}}Ethical concerns, Regulations, \\ Data governance\end{tabular} \\ \midrule 
\textbf{Data modalities} & Uni-modal & Multi-modal (Max two) & Multi-modal (2 or more) \\ \midrule
\textbf{Training time} & Few hours to days & Few days to weeks & Few weeks to months \\ \midrule
\textbf{Dataset size} & 100 to 10k samples & 10k to 100k samples & More than 100k samples \\ \midrule
\textbf{Compute resources} & Cloud & Cloud and Moderate GPUs & In-house high-end GPUs and TPUs \\ \midrule
\textbf{Optimization} & Quantization & \begin{tabular}[c]{@{}l@{}}PEFT techniques, \\ Distillation, Pruning\end{tabular} &
  Prompt compression techniques \\ \midrule
\textbf{Languages} & Monolingual & Monolingual & Multi-lingual/Cross-lingual \\ \midrule 
\textbf{Ethical complexity} & Low & Moderate to high & High to very high \\ \midrule
\textbf{Type and size} & Open-source \textless{}= 3B & Open-source $\sim$7B & Any open-source model \\ \bottomrule
\end{tabular}%
}
\caption{A suggestive guide to various industries to maximize the utilization of LLMs for NLG applications.}
\vspace{-5mm}
\label{tab:guide}
\end{table*}
\subsection{Ethical Concerns} 
The most common ethical challenges with LLMs are violation of the model license, model theft, copyright infringement, producing harmful content, and trustworthiness \cite{paper_015} and following are the \colorbox{gray!20}{\textbf{potential solutions}}. \\
\textbf{Protecting LLMs.} Watermarking techniques \cite{paper_017} are essential for copy-right protection of industrial LLMs, aim to minimize the adverse impact on the original LLM. \\
\textbf{Enhancing creativity.} AI models should enhance, not replace, human creativity by generating new ideas and insights \cite{paper_103}. \\
\textbf{Fairness in data visualization.} Interactive data visualization can help detect and address hidden biases \cite{paper_016}. \\
\textbf{Linking models.} Techniques such as LLM Attribution \cite{paper_015} link fine-tuned models to their pre-trained versions. \\
\textbf{Protecting integrity.} Guardrails such as NeMo \cite{paper_066}, LangKit\footnote{\url{https://docs.whylabs.ai/docs/langkit-api/}}, and TrustLLM \cite{sun2024trustllm} help to maintain LLM integrity by preventing biased or inaccurate outputs.\\
 Addressing these challenges requires a combination of technical expertise, ethical considerations, and further research efforts. In Figure~\ref{fig:life_cycle_fig}, we categorize each paper (total of 68) based on its application life cycle and observed that, due to the above-mentioned pitfalls, more than 70\% of LLM-based studies are still in the conceptual phase.

\section{Maximizing LLM Utilization Across Industries}
\vspace{-2mm}
This section offers a suggestive guide to various industries to maximize the utilization of LLMs for Natural Language Generation (NLG) applications (\textbf{RQ3}). As shown in Table~\ref{tab:guide}, our suggestions are tailored to various industries, considering their distinct goals, resources, and workforce capabilities. The recommendations for small and medium-sized industries equally apply to large-scale industries.\\
\textbf{1) Small-scale industries} such as startups with less than 100 employees need to optimize the use of LLMs within constraints of limited computational resources and workforce. These industries should emphasize prompt engineering and transfer learning techniques to utilize robust small LLMs with up to 3 billion parameters with permissive licenses. Further, these industries should focus on monolingual tasks and actively perform the inference on a few hundred samples. To reduce the inference duration, these industries should opt for optimization techniques such as quantization. Moreover, these industries encounter challenges such as potential reductions in model accuracy, costs, and need for technical expertise. Some of these can be addressed by partnering with AI consulting firms.\\
\textbf{2) Medium-scale industries} up to 1000 employees should focus on utilizing the RAG-based pipelines and domain-specific parameter efficient fine-tuning and distillation techniques for LLMs up to 7B parameters. Additionally, these industries can develop domain-specific adapters to enhance LLMs' performance on specific tasks. These industries can explore moderate multi-modal (text + vision) tasks. Additionally, the key challenges for medium-scale industries are scalability and domain adoption. \\
\textbf{3) Large scale industries} such as MNCs should focus on continuous pre-training of LLMs while ensuring compliance with regulatory requirements. These industries can leverage LLMs effectively across multi-lingual, cross-lingual, and multi-modal generation tasks. Training such models can take from a few weeks to months, which requires high-quality data and huge compute as well. These industries should focus on establishing several collaborative tools and frameworks to maximize LLM utilization.
For all industries, we recommend using open-source models with appropriate licenses to address ethical concerns and comply with LLM regulatory guidelines. Additionally, robust testing and validation protocols are essential to meet industry standards. Fostering strong collaborations and knowledge sharing between industry and academia is crucial for advancing responsible LLM development and deployment.
\section{Conclusions}
 This study delves into the utilization of Large Language Models (LLMs) through an industrial lens, with a specific focus on identifying roadblocks to their adoption. It meticulously examines various pitfalls and provides potential solutions. Moreover, this study offers a guide to organizations of all sizes to maximize the utilization of LLMs for industrial use cases. By identifying pitfalls and suggesting potential directions, the study offers a strategic road-map for optimizing LLM effectiveness in industrial operations. 
\section{Limitations}
Our study has the following limitations. \\
\textbf{Scope.} To provide a practical guide to various industries, we restrict our scope to only Natural Language Generation (NLG) applications. Prospective works should focus on providing an extensive guide to various other tasks as well.\\
\textbf{Coverage.} With the rapid development of LLMs and the voluminous research in this field, it's not feasible to comprehensively cover all the papers. Recognizing this, our survey has focused specifically on industry-related papers. This allowed us to delve deeper and gain an understanding of the unique requirements and challenges faced within industrial applications of LLMs.\\ 
\textbf{Confidentiality.} Due to the confidential nature of the industrial applications not many details were available for specific scenarios or challenges. Hence, we only focused on providing recommendations/insights that can be applicable to a broad range of industrial applications.
\section{Ethics Statement}
To our knowledge, this study presents minimal ethical concerns. However, to maintain transparency, we provide a detailed analysis of all 68 papers present in the survey in Appendix Section~\ref{sec:check_list_survey}. Each paper is reviewed by at least three individuals to validate its claims and findings. We conduct the industrial case study, following the guidelines outlined by the ACL ethics review policy \footnote{\url{https://aclrollingreview.org/ethicsreviewertutorial}}, thereby Ethics Review Boards (ERB) approval is not necessary. It's important to note that our research involving human subjects does not entail the collection of any medical or sensitive information from the users.

\bibliography{custom}

\newpage
\appendix

\section{Survey Papers Selection Criteria}
\label{sec:papers_selection}
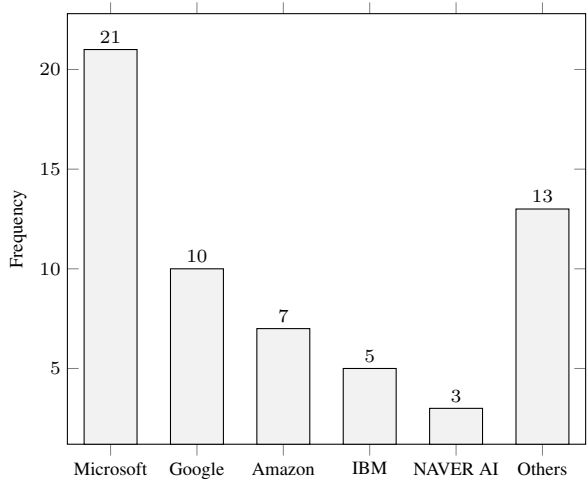
\begin{figure}
\centering\scriptsize
\begin{tikzpicture}
  \begin{axis}[ ybar, bar width=0.7cm, ylabel={Frequency},
    symbolic x coords={Microsoft, Google, Amazon, IBM, NAVER AI, Others},
    xtick=data, nodes near coords, nodes near coords align={vertical},
    ]
    \addplot[fill=black!5] coordinates {
      (Microsoft, 21)
      (Google, 10)
      (Amazon, 7)
      (IBM, 5)
      (NAVER AI, 3)
      (Others, 13)
    };
  \end{axis}
\end{tikzpicture}
\caption{Distribution of research papers from industrial organizations. Others include Apple, Sony, Alibaba, Allen Inst for AI, JP Morgan, Nvidia, Adobe.}
\label{fig: organizations_plot}
\end{figure}

We used keywords such as ``large language models”, ``LLM” and ``LLMs for industrial applications” for selecting the relevant papers. We selected the majority of papers from the reputed databases including the ACL Anthology\footnote{\url{https://aclanthology.org/}}, ACM Digital library\footnote{\url{https://dl.acm.org/}}, Google Scholar\footnote{\url{https://scholar.google.com/}}, which are known for hosting peer-reviewed articles that meet high academic standards. Subsequently, we finalize suitable research papers for the survey based on the following criteria.
\begin{table}[htp]
\centering
\begin{tabular}{lc}
\toprule
\textbf{Criteria} & \textbf{Number of papers} \\ \midrule
arXiv version & 37 \\ 
Non organizational papers & 10 \\ 
Not related to application & 6 \\ 
\textbf{Relevant} & \textbf{68} \\ \midrule
Total & 121 \\ \bottomrule
\end{tabular}
\caption{Survey papers filtration criteria.}
\label{tab:selction_criteria}
\end{table}
\begin{itemize}
    \item The paper should be a peer-reviewed and published version.
    \item At least one of the paper's authors should be from the industry.
    \item Paper should use at least one or more LLM.
    \item The paper should report at least one real-world application using LLM(s).
\end{itemize}
\textbf{Necessary Concessions:} We believe that having at least one author from the industry brought the following advantages.
\begin{itemize}
    \item We found that considering papers with only researchers from industry led to very few research papers. Also, in recent times, collaboration between academia and industry has rightfully expanded resulting in more practical and applicable research works.
    \item Also, they brought practical perspectives that were grounded in real-world applications and challenges.
\end{itemize}
In total, we have collected 121 research papers, and out of them, we have discarded 53 that do not fall under one or more above-mentioned criteria as mentioned in Table~\ref{tab:selction_criteria}. We have omitted 40 papers because they are not peer-reviewed and 10 more papers came from the non-organizations typically submitted by academic labs/universities. Moreover, we have discarded six papers, which did not discuss any industrial application. After applying the filtering criteria we left with 68 relevant papers. This distribution of the list of papers from various industrial organizations is mentioned in Figure~\ref{fig: organizations_plot}.   

\section{Industrial Case Study Details}
\label{sec:case_study}
\begin{table*}[htb]
\begin{tabular}{|p{\textwidth}|}
\hline
\begin{enumerate}
    \item Participant level of expertise in LLMs?
    \begin{itemize}
        \item[\scalebox{1.1}{$\square$}] Beginner 
        \item[\scalebox{1.1}{$\square$}] Intermediate
        \item[\scalebox{1.1}{$\square$}] Proficient
        \item[\scalebox{1.1}{$\square$}] Expert
        \item[\scalebox{1.1}{$\square$}] NA
    \end{itemize}
   \item Application Domain
   \begin{itemize}
       \item[\scalebox{1.1}{$\square$}] Healthcare
       \item[\scalebox{1.1}{$\square$}] Banking
       \item[\scalebox{1.1}{$\square$}] Financial
       \item[\scalebox{1.1}{$\square$}] Retail
       \item[\scalebox{1.1}{$\square$}] Security
       \item[\scalebox{1.1}{$\square$}] Privacy
       \item[\scalebox{1.1}{$\square$}] Legal
       \item[\scalebox{1.1}{$\square$}] Marketing \& Advertising
       \item[\scalebox{1.1}{$\square$}] Education
       \item[\scalebox{1.1}{$\square$}] Media and entertainment
       \item[\scalebox{1.1}{$\square$}] Human Resources(HR)
       \item[\scalebox{1.1}{$\square$}] eCommerce
       \item[\scalebox{1.1}{$\square$}] Other: \rule{2cm}{0.4pt}
   \end{itemize}
   \item What is the name of the task that LLM(s) performs in your project?
   \item Type of data used?
   \begin{itemize}
        \item[\scalebox{1.1}{$\square$}] Tabular
        \item[\scalebox{1.1}{$\square$}] Image
        \item[\scalebox{1.1}{$\square$}] Video
        \item[\scalebox{1.1}{$\square$}] Audio
        \item[\scalebox{1.1}{$\square$}] Text
        \item[\scalebox{1.1}{$\square$}] More than one modality
        \item[\scalebox{1.1}{$\square$}] Other: \rule{2cm}{0.4pt}      
   \end{itemize}
   \item How are the LLMs used?
   \begin{itemize}
       \item[\scalebox{1.1}{$\square$}] Fine-tuning
       \item[\scalebox{1.1}{$\square$}] Zero-shot
       \item[\scalebox{1.1}{$\square$}] In-context learning
       \item[\scalebox{1.1}{$\square$}] Other: \rule{2cm}{0.4pt}
   \end{itemize}
   
\end{enumerate} \\
\hline
\end{tabular}
\caption{Questionnaire for industrial case study: Part 1}
\label{tab:survey_q1}
\end{table*}

\begin{table*}[htb]
\begin{tabular}{|p{\linewidth}|}
\hline
\begin{enumerate}
   \setcounter{enumi}{5}
   \item Did you consider any of the following Trust attributes or guard rails while designing/implementing the LLM-based solution?
   \begin{itemize}
       \item[\scalebox{1.1}{$\square$}] Security
       \item[\scalebox{1.1}{$\square$}] Robustness
       \item[\scalebox{1.1}{$\square$}] Privacy
       \item[\scalebox{1.1}{$\square$}] Bias \& Fairness
       \item[\scalebox{1.1}{$\square$}] Interpretability or Explainability
       \item[\scalebox{1.1}{$\square$}] Toxicity
       \item[\scalebox{1.1}{$\square$}] Hallucination
       \item[\scalebox{1.1}{$\square$}] None
       \item[\scalebox{1.1}{$\square$}] Other: \rule{2cm}{0.4pt}
   \end{itemize}
   \item Name of the LLMs being used?
   \begin{itemize}
       \item[\scalebox{1.1}{$\square$}] LLaMA
       \item[\scalebox{1.1}{$\square$}] LLaMA-2
       \item[\scalebox{1.1}{$\square$}] Falcon
       \item[\scalebox{1.1}{$\square$}] Mistral
       \item[\scalebox{1.1}{$\square$}] GPT3.5 (ChatGPT)
       \item[\scalebox{1.1}{$\square$}] GPT4
       \item[\scalebox{1.1}{$\square$}] MPT
       \item[\scalebox{1.1}{$\square$}] Meta OPT
       \item[\scalebox{1.1}{$\square$}] Bard
       \item[\scalebox{1.1}{$\square$}] PaLM
       \item[\scalebox{1.1}{$\square$}] Pythia
       \item[\scalebox{1.1}{$\square$}] Cerebras-GPT
       \item[\scalebox{1.1}{$\square$}] NA
       \item[\scalebox{1.1}{$\square$}] Other: \rule{2cm}{0.4pt}
   \end{itemize}
   \item What are the risks associated with the LLMs being used in your project?
   \begin{itemize}
       \item[\scalebox{1.1}{$\square$}] Security and Safety
       \item[\scalebox{1.1}{$\square$}] Reputation
       \item[\scalebox{1.1}{$\square$}] Quality of service
       \item[\scalebox{1.1}{$\square$}] Revenue
       \item[\scalebox{1.1}{$\square$}] License
       \item[\scalebox{1.1}{$\square$}] NA
       \item[\scalebox{1.1}{$\square$}] Other: \rule{2cm}{0.4pt}
   \end{itemize}
\end{enumerate} \\
\hline
\end{tabular}
\caption{Questionnaire for industrial case study: Part 2}
\label{tab:survey_q2}
\end{table*}


We have created a questionnaire to conduct the industrial case study as shown in Fig~\ref{tab:survey_q1}.

\section{Survey Papers Checklist}
\label{sec:check_list_survey}
This paper provides a review of 68 papers and for each paper, we reported 22 features as mentioned in Table~\ref{tab:check_list}. We briefly describe each feature in the master table for better understanding.
\begin{itemize}
    \setlength{\itemsep}{0pt}
\item \textit{Paper:} Citation of the paper.
\item \textit{Venue:} The venue where the paper was published.
\item \textit{Year:} Year of paper publication.
\item \textit{LLM name:} Names of the LLMs used in the paper.
\item \textit{Organization:} Name of the industrial organization involved in the work.
\item \textit{Domain:} Domain information of the application in the paper.
\item \textit{Application:} The type of application under which the work was categorized into.
\item \textit{Use case:} The information of how the paper leverages an LLM in a specific scenario or a task. 
\item \textit{Dataset Name:} Datasets used by the paper for modeling and evaluation. 
\item \textit{Prompting Strategy:} Prompting strategies used in the paper. 
\item \textit{Evaluation metrics:} Details of the evaluation metrics used in the paper. 
\item \textit{Application life cycle:} Information of application's life cycle stage.
\item \textit{GitHub:} Link to the GitHub repository, if any, that was published in the paper.
\item \textit{License:} This field indicates if the paper contains license-related information.
\item \textit{Privacy:}This field indicates if the paper contains privacy-related information.
\item \textit{Use cases:} This field indicates if the paper mentions a use case or not.
\item \textit{Limitations:} Major limitations of the paper, if any.
\end{itemize}
\useunder{\uline}{\ul}{}
\begin{sidewaystable}[htb]
\vspace{50ex}
\centering\scriptsize
\resizebox{\textwidth}{!}{%
\begin{tabular}{cllrllllllllllllll}
\toprule
 & \multicolumn{1}{l}{\textbf{Paper}} & \multicolumn{1}{c}{\textbf{Venue}} & \multicolumn{1}{c}{\textbf{Year}} & \multicolumn{1}{c}{\textbf{LLMs used}} & \multicolumn{1}{c}{\textbf{Organization}} & \multicolumn{1}{c}{\textbf{Domain}} & \multicolumn{1}{c}{\textbf{Application}} & \multicolumn{1}{c}{\textbf{Use case}} & \multicolumn{1}{c}{\textbf{Dataset Name}} & \multicolumn{1}{c}{\textbf{Prompting strategy}} & \multicolumn{1}{c}{\textbf{Evaluation metrics}} & \multicolumn{1}{c}{\textbf{Application life cycle}} & \multicolumn{1}{c}{\textbf{Github}} & \multicolumn{1}{c}{\textbf{License}} & \multicolumn{1}{c}{\textbf{Privacy}} & \multicolumn{1}{c}{\textbf{Use cases}} & \multicolumn{1}{c}{\textbf{Limitations}}   \\ \midrule
1 & \citet{li-etal-2023-chatgpt} & EMNLP Industry Track & 2023 & \begin{tabular}[c]{@{}l@{}}ChatGPT, GPT-4, BloombergGPT, \\ GPT-NeoX, OPT66B, BLOOM176B, \\ FinBERT\end{tabular} & J.P. Morgan AI Research & Financial & Analytics & Financial text analysis & \begin{tabular}[c]{@{}l@{}}FPB/FiQA/TweetFinSent, Headline, \\ NER, REFinD, FinQA/ConvFinQA\end{tabular} & Zero-shot, Few-shot and CoT & Accuracy, F1 Score & Conceptualization/PoC & NA & Yes & NA & Yes & NA   \\
2 & \citet{chen2023empowering} & EuroSys & 2024 & GPT-3.5, GPT-4 & Microsoft & Fault diagnosis & Cloud management & Cloud incident root cause analysis & \begin{tabular}[c]{@{}l@{}}653 incidents from Microsoft's transport service\end{tabular} & Zero-shot & Micro and Macro F1-score & Conceptualization/PoC & NA & NA & NA & NA & Effectivness of the method depends on incident monitors/alerts. \\
3 & \citet{paper_035} & FMDM@NeurIPS & 2023 & GPT-4-32k & Microsoft research & Software & Code generation & Automate repository level code planning tasks & Proprietary & Zero-shot & Block metrics, Edit metrics & Conceptualization/PoC & NA & NA & NA & NA & \begin{tabular}[c]{@{}l@{}}Dynamic languages may not be ideal for a coded plan approach\end{tabular}   \\
4 & \citet{paper_036}  & HotNet's & 2023 & \begin{tabular}[c]{@{}l@{}}GPT-4, GPT-3, Text-davinci-003,\\ Bard\end{tabular} & Microsoft research & Communication & Code generation & Code generation for graph manipulation tasks & Public code repositories & Zero-shot & Accuracy & Conceptualization/PoC &  {\ul \href{https://github.com/microsoft/NeMoEval}{Link}} & Yes & NA & NA & \begin{tabular}[c]{@{}l@{}} High quality domain specific code synthesis is still an open challenge\end{tabular}   \\
5 & \citet{paper_053}  & ACL & 2023 & \begin{tabular}[c]{@{}l@{}}CodeGen-350M,CodeGen-2B,\\ CodeGen-6B,CodeGen-16B,\end{tabular} & AWS AI Labs & Software & Code generation & Static code analysis for completion & Function completion dataset & None & \begin{tabular}[c]{@{}l@{}}Percentages of AST errors, \\ Undefine variable,  unused variables etc.\end{tabular} & Conceptualization/PoC & NA & NA & No & Yes & \begin{tabular}[c]{@{}l@{}} Cross-file context based broader categorization of errors \\ was not employed\end{tabular}   \\
6 & \citet{paper_082} & UIST & 2023 & \begin{tabular}[c]{@{}l@{}}GPT-3.5 \\  Legacy (text-davinci-003),\\  Legacy (code-davinci-002),\\  Legacy (text-davinci-edit-001)\end{tabular} & NA & Software & Code generation & Webpage customization using LLMs & NA & Fewshot (In-context learning) & NA & Prototype & NA & No & No & Yes & \begin{tabular}[c]{@{}l@{}} Poor performance on complex website customization\end{tabular}   \\
7 & \citet{paper_091} & ICER & 2023 & GPT-3.5, GPT-4 & Microsoft & Software & Code generation & \begin{tabular}[c]{@{}l@{}}LLM tutor for programming education\end{tabular} & NA & Zero-shot & Match & Conceptualization/PoC & NA & NA & NA & Yes & \begin{tabular}[c]{@{}l@{}}Limited to python language and introductory educational content\end{tabular}   \\
8 & \citet{paper_093} & ESEC/FSE & 2023 & CODEX, CODEt5 & Microsoft & Software & Code generation & \begin{tabular}[c]{@{}l@{}}Prediction of code edits using LLMs\end{tabular} & C3PO, Overwatch & Fewshot (In-context learning) & Exact match of the code segment & Conceptualization/PoC & NA & NA & NA & Yes & \begin{tabular}[c]{@{}l@{}} Proposed approach may fail due to LLMs hallucination \\and context length requirements\end{tabular}   \\
9 & \citet{paper_100} & NeurIPS & 2023 & CODEGEN, INCODER & AWS & Software & Code generation & \begin{tabular}[c]{@{}l@{}}Context based code completion\end{tabular} & Buggy-HumanEval, Buggy-FixEval & Few-shot and CoT & pass@k & Conceptualization/PoC &  {\ul \href{https://github.com/amazon-science/buggy-code-completion}{Link}} & Yes & No & Yes & \begin{tabular}[c]{@{}l@{}}Proposed method may not be aligned to general software development \\ setting as buggy datasets are based on programming contest submissions\end{tabular}   \\
10 & \citet{paper_106} & ICLR & 2023 & Decoder-only transformer models & AWS AI Labs & Software & Code generation & Evaluation LLMs on multilingual programming datasets & MBXP, Multilingual HumanEval, MathQA-X & Zero-shot and Few-shot & pass@k scores & Conceptualization/PoC & {\ul \href{https://github.com/amazon-science/mxeval}{Link}} & Yes & NA & Yes & Lack of language specific evaluations  \\ 
11 & \citet{paper_057} & CHI & 2023 & HyperCLOVA & NAVER AI Lab, NAVER CLOUD & Healthcare & Conversational & Voice assistant & NA & Zero-shot & NA & Deployment & {\ul \href{https://guide.ncloud-docs.com/docs/en/clovacarecall-overview}{Link}} & NA & No & Yes & \begin{tabular}[c]{@{}l@{}}Skewed age distribution in pilot subjects and pilot deployment time\end{tabular}   \\
12 & \citet{paper_059} & CHI & 2023 & NA & Google Research & Accessibility & Conversational & Evaluation of LLMs as a tool for AAC users & Proprietary & None & NA & Conceptualization/PoC & NA & NA & NA & Yes & NA   \\
13 & \citet{paper_073} & FAccT & 2023 & LaMDA & Google Research & Accessibility & Conversational & Chatbot & NA & None & Human evaluation & NA & NA & Yes & No & Yes & Limited diversity of identity within individual focus groups   \\
14 & \citet{paper_049} & NeurIPS & 2023 & \begin{tabular}[c]{@{}l@{}}Codex, GPT-3.5, GPT-3.5-chat,\\ GPT-4\end{tabular} & Google & Generic & Data Generation & \begin{tabular}[c]{@{}l@{}}Visual planning for text-to-image generation\end{tabular} & NSR-1K, 3D-FRONT & Fewshot (In-context learning) & \begin{tabular}[c]{@{}l@{}}CLIP cosine similarity,\\GLIP accuracy, \\ Attribute binding Accuracy, \\ KL divergence\end{tabular} & Conceptualization/PoC &  {\ul \href{https://github.com/weixi-feng/LayoutGPT}{Link}} & Yes & NA & Yes & \begin{tabular}[c]{@{}l@{}} Generation of overly dense layouts and unusual sized bounding boxes\end{tabular}   \\
15 & \citet{paper_064} & ACL & 2023 & \begin{tabular}[c]{@{}l@{}}Used existing LLMs from Hugginhface, \\ openAI, Azure, Anthropic, Cohere\end{tabular} & Deepset GMBH & Generic & Data Generation & Generation of labeled training data & IMDB, MRPC, SNLI, TREC-6, SQUAD & Zero-shot and Few-shot & F1 score & Conceptualization/PoC & {\ul \href{https://github.com/flairNLP/fabricator}{Link}} & No & No & Yes & \begin{tabular}[c]{@{}l@{}}Evaluation cover only subset of commonly encountered tasks\end{tabular}   \\
16 & \citet{paper_008} & EMNLP Industry Track & 2023 & GPT-4, LLaMA & Amazon & Financial & Forecasting & \begin{tabular}[c]{@{}l@{}}Explainable financial time series forecasting\end{tabular} & \begin{tabular}[c]{@{}l@{}}Stock price data, Company profile data, \\  Finance/Economy News Data\end{tabular} & Zero-shot and Few-shot & \begin{tabular}[c]{@{}l@{}}Binary Precision, Bin Precision, \\ MSE, ROUGE-1,2\end{tabular} & Conceptualization/PoC & NA & Yes & Yes & NA & \begin{tabular}[c]{@{}l@{}}Generalization to other types of financial temporal data unexplored\end{tabular}   \\
17 & \citet{paper_025} & NeurIPS & 2023 & \begin{tabular}[c]{@{}l@{}}Language model for mixed reality (LLMR)\\ Dall.E-2\\ GPT-4\end{tabular} & Microsoft, Microsoft Research & Generic & Frameworks & \begin{tabular}[c]{@{}l@{}}Generation of interactive 3D objects\end{tabular} & NA & NA & \begin{tabular}[c]{@{}l@{}}Error rate, Average generation time.\end{tabular} & Development & NA & NA & NA & Yes & \begin{tabular}[c]{@{}l@{}}For complex tasks manual code editing might be still necessary\end{tabular}   \\
18 & \citet{paper_046}  & CHI & 2023 & PaLM & Google Research & NLP & Frameworks & \begin{tabular}[c]{@{}l@{}}Conversational Interaction \\ with Mobile UI\end{tabular} & \begin{tabular}[c]{@{}l@{}}PixelHelp , AndroidHowTo, \\ Rico, Screen2Words,\end{tabular} & Zero-shot and Few-shot & \begin{tabular}[c]{@{}l@{}}Grammar Correctness, UI Relevance,\\ Question Coverage, BLEU, CIDEr,\\ ROUGE-L, and METEOR, Micro-F1\end{tabular} & Conceptualization/PoC & {\ul \href{https://github.com/google-research/google-research/tree/master/llm4mobile}{Link}} & Yes & NA & Yes & \begin{tabular}[c]{@{}l@{}} Fails to handle generation of incorrect or irrelevant information\end{tabular}   \\
19 & \citet{paper_068} & \begin{tabular}[c]{@{}l@{}}EMNLP System \\ demonstrations\end{tabular} & 2023 & \begin{tabular}[c]{@{}l@{}}Llama-2 Chat (13B, 70B), \\ WizardVicunaLM-13B, \\ Vicuna\end{tabular} & Kioxia Corporation & Generic & Frameworks & \begin{tabular}[c]{@{}l@{}}Framework for knowledge intensive tasks\end{tabular} & KILT Benchmark & NA & \begin{tabular}[c]{@{}l@{}}Exact Match, F1, Accuracy, RL, \\ R-precision\end{tabular} & Deployment & {\ul \href{https://github.com/yhoshi3/RaLLe}{Link}} & Yes & NA & Yes & Falls behind specialized RAG models on KILT tasks  \\
20 & \citet{paper_074} & CoRL & 2023 & LLaMA-13b & Google AI & Robotics & Frameworks & \begin{tabular}[c]{@{}l@{}}LLM guided skill chaining\end{tabular} & ALFRED & Zero-shot & NA & Development & NA & NA & No & Yes & \begin{tabular}[c]{@{}l@{}} Greedy skill chaining may not be optimal for consistent behaviour generation \end{tabular}   \\
21 & \citet{paper_090} & EMNLP & 2023 & \begin{tabular}[c]{@{}l@{}}GPT-3.5-turbo, text-davinci-003, \\ GPT-4-32k\end{tabular} & Microsoft health futures & Healthcare & Frameworks & \begin{tabular}[c]{@{}l@{}}Evaluation of GPT-4 on understanding and \\ generation of radiology tasks\end{tabular} & \begin{tabular}[c]{@{}l@{}}MS-CXR-T, RadNLI, Chest ImaGenome, \\ MIMIC, Open-i\end{tabular} & \begin{tabular}[c]{@{}l@{}}Zero-shot, Few-shot, One-shot, \\ Many-shot, CoT\end{tabular} & \begin{tabular}[c]{@{}l@{}}macro F1, micro F1,  RougeL, \\ CheXbert score\end{tabular} & Deployment & NA & NA & Yes & Yes & \begin{tabular}[c]{@{}l@{}} Qualitative evaluation of the findings on summarization task is \\limited to a single radiologist \end{tabular}   \\
22 & \citet{jiang-etal-2023-llmlingua} & EMNLP & 2023 & GPT-3.5-Turbo-0301 and Claude-v1.3 & Microsoft & NA & None & \begin{tabular}[c]{@{}l@{}}Prompt compression for higher inference speed\end{tabular} & \begin{tabular}[c]{@{}l@{}}GSM8K, BBH, ShareGPT, \\ Arxiv-March23\end{tabular} & Zero-shot & BLEU, ROUGE, BERTScore & Conceptualization/PoC & {\ul \href{https://github.com/microsoft/LLMLingua}{Link}} & Yes & NA & Yes & \begin{tabular}[c]{@{}l@{}} Performance drops when compression of prompts go beyond 25\%\end{tabular}   \\
23 & \citet{paper_003} & EMNLP Industry Track & 2023 & GPT-4, GPT3.5, LLaMA-2 & Microsoft & NLP & Question-answering & \begin{tabular}[c]{@{}l@{}}Domain specific industrial QA\end{tabular} & MSQA & Zero-shot & \begin{tabular}[c]{@{}l@{}}BLEU, ROUGE, METEOR, BERTScore, \\ F1, Key-word/Span-Hit-Rate (KHR), \\ Can-Answer-Rate (CAR), LLM-based Metrics\end{tabular} & Conceptualization/PoC & NA & NA & NA & NA & \begin{tabular}[c]{@{}l@{}}Only works with English data \end{tabular}   \\
24 & \citet{paper_080} & ICAIF & 2023 & GPT-3.5-turbo & J. P. Morgan AI Research & Finance & Question-answering & \begin{tabular}[c]{@{}l@{}}Dynamic workflow generation\end{tabular} & \begin{tabular}[c]{@{}l@{}}NCEN-QA, NCEN-QA-Easy, \\ NCEN-QA-Intermediate, NCEN-QA-Hard\end{tabular} & Zero-shot & NA & Prototype & NA & NA & Yes & Yes & NA   \\
25 & \citet{paper_102} & ICCV & 2023 & GPT-3 & Microsoft, Allen Institute of AI & Vision & Question-answering & \begin{tabular}[c]{@{}l@{}}Natural language based question-aware caption model\end{tabular} & COCO, OK-VQA, A-OKVQA, WebQA & Fewshot (In-context learning) & Accuracy & Conceptualization/PoC &  {\ul \href{https://github.com/nerfies/nerfies.github.io}{Link}} & NA & No & Mix & \begin{tabular}[c]{@{}l@{}}Focuses only on knowledge-based VQA tasks\end{tabular}   \\
26 & \citet{paper_107} & NeurIPS & 2023 & \begin{tabular}[c]{@{}l@{}}GPT-3.5 turbo, GPT-4, \\ ChatGLM, LLaMA, \\ Vicuna, Alpaca\end{tabular} & Alibaba Group, Ant Group & Medical & Question-answering & \begin{tabular}[c]{@{}l@{}}Medical domain QA\end{tabular} & CMExam & Fewshot (In-context learning) & accuracy, weighted F1, BLEU, ROUGE & Conceptualization/PoC & {\ul \href{https://github.com/williamliujl/CMExam/tree/main}{Link}} & Yes & Yes & Yes & \begin{tabular}[c]{@{}l@{}}Excluding non-textual questions might introduce unexpected bias\end{tabular}   \\
27 & \citet{paper_009}  & EMNLP Industry Track & 2023 & text-davinci-002, PaLM & Microsoft & NLP & Reasoning & Mathematical Reasoning & MultiArith dataset & Zero-shot, Few-shot and CoT & Accuracy & Conceptualization/PoC & NA & Yes & NA & NA & \begin{tabular}[c]{@{}l@{}}Non trivial probability of producing incorrect results \\ using algebraic and pythonic expressions\end{tabular}   \\
28 & \citet{paper_037} & NeurIPS & 2023 & GPT-3.5-turbo , GPT-4 & Microsoft research & Generic & Reasoning & Multi-modal knowledge intensive reasoning tasks & ScienceQA, TabMWP & Zero-shot and CoT & Accuracy & Conceptualization/PoC &  {\ul \href{https://github.com/lupantech/chameleon-llm}{Link}} & Yes & NA & NA & Computationally expensive for complex tasks   \\
29 & \citet{paper_051} & CIKM & 2023 & \begin{tabular}[c]{@{}l@{}}GPT-Neo-1.3B, \\ GPT-Neo-2.7B, \\ GPT-J-6B,\\ Falcon-7B-Instruct\end{tabular} & Amazon Alexa AI & Political, Education & Reasoning & \begin{tabular}[c]{@{}l@{}}Steerability of LLM based on persona\end{tabular} & OpinionQA & None & \begin{tabular}[c]{@{}l@{}}User study\end{tabular} & Conceptualization/PoC & NA & NA & No & Yes & Complex personas may not be possible   \\
30 & \citet{paper_108} & SIGIR & 2023 & CODEX & Alibaba Group & Generic & Reasoning & \begin{tabular}[c]{@{}l@{}}Reasoning on large tables based on textual prompts\end{tabular} & TabFact, WikiTableQuestion, FetaQA & Fewshot (In-context learning) & \begin{tabular}[c]{@{}l@{}}binary classification accuracy,\\ denotation accuracy, BLEU, \\ ROUGE-1, ROUGE-2 and ROUGE-L\end{tabular} & Conceptualization/PoC & NA & Yes & Yes & Yes & NA  \\
31 & \citet{paper_118} & ICML Workshop & 2023 & GPT-3.5, GPT-4 & Microsoft Research & Generic & Reasoning & \begin{tabular}[c]{@{}l@{}}LLM based causal QA system\end{tabular} & NA & Text completion & NA & Conceptualization/PoC & NA & NA & NA & Yes & Performance decreases with increased context length \\
32 & \citet{paper_048} & RecSys & 2023 & PaLM & Google & Retail & Recommended systems & \begin{tabular}[c]{@{}l@{}}Visually augmented real-time conversations\end{tabular} & Proprietary & Completion, zero-shot and Few-shot & Mean NDCG & Conceptualization/PoC & NA & No & Yes & Yes & NA   \\
33 & \citet{paper_078} & Advanced robotics & 2023 & Hyperclova & LINE corporation & Generic & Recommender systems & Voice Chatbot & Proprietary & Fewshot (In-context learning) & \begin{tabular}[c]{@{}l@{}}Technical score on Informativeness, \\ Naturalness,  Likability, \\ Satisfaction with dialog\end{tabular} & Testing & NA & NA & No & Yes & \begin{tabular}[c]{@{}l@{}}For low frequency words gives long \\ responses which floods user with information and hallucination\end{tabular}   \\
34 & \citet{paper_114} & ICML Workshop & 2023 & PaLM2 & Walmart Global Tech & Retail & Recommender systems & \begin{tabular}[c]{@{}l@{}}Enhance the capabilities of recommendation \\ systems\end{tabular} & Proprietary & Zero-shot & MRR, NDCG & Conceptualization/PoC & NA & NA & NA & NA & NA   \\
35 & \citet{paper_014} & NeurIPS & 2022 & CLIP & \begin{tabular}[c]{@{}l@{}}IBM Research, \\ MIT-IBM AI-Watson Lab\end{tabular} & Vision & Retrieval & \begin{tabular}[c]{@{}l@{}}Evaluation of LLMs on expert tasks for \textbackslash\\ image-to-text and text-to-image retrieval\end{tabular} & FETA & \begin{tabular}[c]{@{}l@{}}Zero-shot, Few-shot, One-shot, \\ Many-shot\end{tabular} & Accuracy & Conceptualization/PoC & NA & NA & NA & NA & \begin{tabular}[c]{@{}l@{}}FETA contains only a small subset of available \\ technical documents for different expert V\&L data domains\end{tabular}   \\
36 & \citet{paper_042} & ICLR & 2023 & InstructGPT & Microsoft Cognitive Service Research & Generic & Retrieval & LLM based retrieval for knowledge-intensive tasks & TriviaQA, WebQ & Zero-shot & Accuracy, F1, ROUGE-L & Conceptualization/PoC & {\ul \href{https://github.com/wyu97/GenRead}{Link}} & NA & Yes & Yes & \begin{tabular}[c]{@{}l@{}}Limited ability to update knowledge to new domains\end{tabular}   \\
37 & \citet{paper_044} & EMNLP & 2023 & \begin{tabular}[c]{@{}l@{}}Text-davinci-001, Text-davinci-003, \\ GPT-4, Babbage, curie\end{tabular} & Microsoft Research & Generic & Retrieval & Query expansion based retrieval systems & MS-MARCO, TREC DL 2019 & Fewshot & MRR, nDCG & Conceptualization/PoC &  & NA & NA & NA & Efficiency of retrieval system   \\
38 & \citet{paper_047} & CHI & 2023 & GPT3 & Google Research & Generic & Retrieval & \begin{tabular}[c]{@{}l@{}}Augmenting video conferencing \\ with visual captions\end{tabular} & VC 1.5K & Zero-shot & User study & Deployment & {\ul \href{https://github.com/google/archat}{Link}} & Yes & Yes & Yes & \begin{tabular}[c]{@{}l@{}} Visual captions in conversations should have a threshold\\  to filter out potentially distracting or inappropriate content \end{tabular}  \\
39 & \citet{paper_052} & NeurIPS & 2023 & \begin{tabular}[c]{@{}l@{}}GPT2-S (117M), \\ GPT2-L (774M) {[}29{]},\\ OpenLLaMA-7B (7B)\end{tabular} & \begin{tabular}[c]{@{}l@{}}AWS GAIIC,\\ AWS AI\end{tabular} & Healthcare & Retrieval & Writing radiology reports from medical images & MIMIC-CXR & None & \begin{tabular}[c]{@{}l@{}}Factual completeness and correctness\\ F1-CXB-14 score, F1-CXB-5, \\ BLEU4, ROUGE-L\end{tabular} & Conceptualization/PoC &  {\ul \href{https://aws.amazon.com/machine-learning/responsible-machine-learning/aws-healthscribe/}{Link}} & NA & No & Yes & \begin{tabular}[c]{@{}l@{}}soft visual prompt doesn’t receive consistent attention, \\ especially when using LLMs.\end{tabular}   \\
40 & \citet{paper_094} & SIGIR & 2023 & text-davinci-003 & Microsoft & Generic & Retrieval & \begin{tabular}[c]{@{}l@{}}Generation of query variants for building \\ test collections and document pool\end{tabular} & UQV100 & One-shot & \begin{tabular}[c]{@{}l@{}} Jaccard Index, RBP, RBO\end{tabular} & Conceptualization/PoC & NA & NA & NA & Yes & \begin{tabular}[c]{@{}l@{}} Small size human-generated data is not \\ sufficient for few-shot prompting\end{tabular}   \\
41 & \citet{paper_001} & EMNLP Industry Track & 2023 & CTI-BERT & IBM T. J. Watson Research Center & Security & Security & Cyber threat intelligence & \begin{tabular}[c]{@{}l@{}}Attack description, Security Textbook,\\  Academic Paper, Security Wiki, \\ Threat reports, Vulnerability\end{tabular} & NA & Micro and Macro F1 Score & Conceptualization/PoC & NA & Yes & Yes & Yes & Pretrained only on English data.   \\
42 & \citet{paper_015}  & ACL & 2023 & \begin{tabular}[c]{@{}l@{}}BERT, GPT, BLOOM, codegen-350M, \\ DialoGPT, DistilGPT2, \\ OPT, GPT-Neo, xlnet-base-cased, \\ multilingual-miniLM-L12-v2\end{tabular} & IBM Research & Generic & Security & \begin{tabular}[c]{@{}l@{}}Tracing back to the origin of fine-tuned models to \\ alleviates the problem of accountability of LLMs\end{tabular} & \begin{tabular}[c]{@{}l@{}}GitHub, The BigScience ROOTS Corpus, \\ CC-100, Reddit, and THEPILE\end{tabular} & NA & F1, ROC & Conceptualization/PoC & NA & Yes & Yes & Yes & \begin{tabular}[c]{@{}l@{}}Considered only a limited number of LLMs for the study.\end{tabular}   \\
43 & \citet{paper_017} & ACL & 2023 & text-embedding-ada-002, BERT & \begin{tabular}[c]{@{}l@{}}Microsoft Research Asia, Sony AI,\\  Micorsoft STC Asia\end{tabular} & Security & Security & \begin{tabular}[c]{@{}l@{}}Copy right protection of EaaS (Embeddings as \\ a Service) LLMs\end{tabular} & SST2, Mind, Enron Spam, AG news & NA & Accuracy, Detection performance & Conceptualization/PoC & NA & NA & NA & NA & NA   \\
44 & \citet{paper_099} & WWW & 2023 & GPT-3, PaLM & Google Research & Finance & Sentiment Analysis & \begin{tabular}[c]{@{}l@{}}Labels generation for financial data\end{tabular} & FiQA-News & Fewshot (In-context learning) & Accuracy & Conceptualization/PoC & NA & NA & No & Yes & -   \\
45 & \citet{paper_010} & EMNLP Industry Track & 2023 & HyperCLOVA (30B and 82B), and GPT-3 & NAVER AI Lab & Generic & Societal Impact & Social bias risk mitigation& KoSBi & NA & F1 Score & Conceptualization/PoC & {\ul \href{https://github.com/naver-ai/korean-safety-benchmarks}{Link}} & Yes & Yes & Yes & \begin{tabular}[c]{@{}l@{}}The performance of the filter models are not very competitive\end{tabular}   \\
46 & \citet{paper_026} & EMNLP Industry Track & 2023 & \begin{tabular}[c]{@{}l@{}}GPT-3.5-turbo-0301, Falcon 40B-instruct, \\ Falcon 7B-instruct, Dolly-v2-12b\end{tabular} & Apple & Security & Societal Impact & \begin{tabular}[c]{@{}l@{}}Comprehensive handling of controversial issues\end{tabular} & DELPHI & Zero-shot & \begin{tabular}[c]{@{}l@{}}Controversy Acknowledgement Rate, \\ Comprehensiveness Answer Rate\end{tabular} & Conceptualization/PoC & {\ul \href{https://github.com/apple/ml-delphi}{Link}} & Yes & Yes & Yes & \begin{tabular}[c]{@{}l@{}}Dataset may not cover all the controversial questions.\\ and may contain expired ground truth controversy labels\end{tabular}   \\
47 & \citet{paper_106} & ICLR & 2023 & Decoder-only transformer model & AWS AI Labs & Software & Code generation & \begin{tabular}[c]{@{}l@{}} Evaluation of LLMs on multilingual \\ programming datasets \end{tabular} & MBXP, Multilingual HumanEval, MathQA-X & Zero-shot and Few-shot & pass@k scores & Conceptualization/PoC & {\ul \href{https://github.com/amazon-science/mxeval}{Link}} & Yes & NA & NA & \begin{tabular}[c]{@{}l@{}}Does not support language-specific functionalities\end{tabular}   \\
48 & \citet{paper_004}  & EMNLP Industry Track & 2023 & \begin{tabular}[c]{@{}l@{}}GPT-4, GPT3.5, PaLM-2, \\ LLaMA-2 13b, 7b\end{tabular} & Dialpad Canada Inc & NLP & Summarization & Business meeting summarization & QMSUM, AMI, ICSI & Zero-shot & ROUGE, BERTScore & Conceptualization/PoC & NA & Yes & Yes & Yes & \begin{tabular}[c]{@{}l@{}}Generalizability to domain-specific datasets is in question \\because only academic datasets were used for testing\end{tabular}   \\
49 & \citet{paper_018} & EMNLP  Industry Track & 2023 & FLAT-T5 & Amazon & NLP & Summarization & Summarization of length product titles & NA & NA & ROUGE, BLEU & Conceptualization/PoC & NA & NA & NA & NA & Do not guarantee inclusion of salient words in the summary \\
50 & \citet{paper_021} & RecSys & 2023 & Alpaca-LoRa & Sony Research India & Retail & Summarization & \begin{tabular}[c]{@{}l@{}}Generation of product descriptions sans webscrapping\end{tabular} & MovieLens, Goodreads Book graph & Fewshot (In-context learning) & \begin{tabular}[c]{@{}l@{}}Hit Rate, Normalized Discount \\ Cumulative Gain (NDCG), \\ Mean Reciprocal Rank (MRR)\end{tabular} & Conceptualization/PoC & NA & NA & NA & NA & \begin{tabular}[c]{@{}l@{}}Generates factually incorrect descriptions\end{tabular}   \\
51 & \citet{paper_098} & ESEC/FSE & 2023 & GPT-3.X & Microsoft & Infrastructure & Summarization & \begin{tabular}[c]{@{}l@{}}Cloud outage management\end{tabular} & Proprietary & NA & BLEU-4, ROUGE-L and METEOR & Deployment & NA & NA & NA & Yes & Evaluation metrics not fully reflect readability and usefulness of outage summary   \\
52 & \citet{paper_103} & CHI In2Writing Workshop & 2023 & NA & Allen Institute of AI & NLP & Summarization & \begin{tabular}[c]{@{}l@{}}Evidence based knowledge generation\end{tabular} & NA & Fewshot (In-context learning) & NA & NA & NA & NA & NA & Yes & NA   \\
53 & \citet{paper_002} & EMNLP Industry Track & 2023 & \begin{tabular}[c]{@{}l@{}}GPT4, TULU, Pythia,  Alpaca,\\  Vicuna, LLaMA-2, GPT-3.5\end{tabular} & Allen Institute for AI & NLP & Table-to-text-generation & \begin{tabular}[c]{@{}l@{}}Evaluation of LLMs on table-to-text generation\end{tabular} & LOTNLG, F2WTQ & Zero-shot and Few-shot & \begin{tabular}[c]{@{}l@{}}BLEU, ROUGE, SP-Acc, NLI-Acc, \\ TAPAS-Acc, TAPEX-Acc, Exact-match, \\ F1 Score, Accuracy\end{tabular} & Conceptualization/PoC & NA & Yes & NA & Yes & NA   \\
54 & \citet{paper_013} & ISWC & 2023 & GPT4, Llama2, FLAN-T5 & IBM Research & NLP & Table-to-text-generation & \begin{tabular}[c]{@{}l@{}}Automation of meta data\\ generation and enrichment\end{tabular} & NA & None & NA & Conceptualization/PoC & NA & NA & NA & NA & NA   \\
55 & \citet{paper_032} & NeurIPS & 2023 & GPT-3.5 (text-davinci-003 endpoint) & Microsoft & Generic & Table-to-text-generation &\begin{tabular}[c]{@{}l@{}} Noise induction to better \\understanding of table structures \end{tabular} & \begin{tabular}[c]{@{}l@{}}AirQuality, HousingData, Diabetes, \\ Wine Testing, Iris, Titanic, \\ and ENB2012\_data\end{tabular} & Zero-shot & F1 score & Conceptualization/PoC & {\ul \href{https://github.com/microsoft/prose}{Link}} & Yes & NA & NA & \begin{tabular}[c]{@{}l@{}}Performance of structural tasks with downstream task such as table \\ question answering remains an open challenge.\end{tabular}   \\
56 & \citet{paper_041} & IEEE Access & 2023 & ChatGPT & Microsoft & Robotics & Task Planning & \begin{tabular}[c]{@{}l@{}}Translating natural language instructions \\ to executable robot actions\end{tabular} & NA & Fewshot & Executability, Correctness & Conceptualization/PoC &  {\ul \href{https://github.com/microsoft/ChatGPT-Robot-Manipulation-Prompts}{Link}} & Yes & Yes & Yes & Only static environment is considered   \\
57 & \citet{paper_007} & EMNLP Industry Track & 2023 & GPT- text-embedding-ada-002 & Kingfisher Labs Ltd , Just Access & Legal & Tool & \begin{tabular}[c]{@{}l@{}}Automatic linkage of judgements to bookmarks \\ in court hearing videos\end{tabular} & UK National Archive & Zero-shot & Mean Average Precision (MAP), Recall & Conceptualization/PoC & NA & Yes & Yes & NA & NA   \\
58 & \citet{paper_050} & CHI Extended Abstract & 2023 & - & Google Research & Generic (HCI) & Tool & input-output interaction , Frame change & NA & Zero-shot & Questionnaire & Conceptualization/PoC & NA & NA & NA & Yes & \begin{tabular}[c]{@{}l@{}}Needs formal evaluation and in-depth analysis on \\how functional prompts effect prototyping process\end{tabular}   \\
59 & \citet{dibia-2023-lida} & ACL & 2023 & NA & Microsoft Research & Generic & Tool & \begin{tabular}[c]{@{}l@{}}Automatic generation of grammar-agnostic \\ visualizations and infographics\end{tabular} & Proprietary & Zero-shot and Few-shot & \begin{tabular}[c]{@{}l@{}}Visualization Error Rate (VER),\\ Self-Evaluated Visualization Quality\\ (SEVQ), code accuracy, \\ data transformation, goal compliance, \\ visualization type, data encoding,\\ and aesthetics\end{tabular} & Prototype & {\ul \href{https://microsoft.github.io/lida/}{Link}} & Yes & No & Yes & \begin{tabular}[c]{@{}l@{}}Code execution step increases computational complexity.\end{tabular}   \\
60 & \citet{paper_072} & ICRA & 2023 & \begin{tabular}[c]{@{}l@{}}text-davinci-*,\\ Codex,\\ GPT3\end{tabular} & Nvidia corporation & Robotics & Tool & Generate programmatic robot instructions using LLMs & NA & Fewshot (In-context learning) & \begin{tabular}[c]{@{}l@{}}Success rate(SR),\\ Goal condition recall(GCR),\\ Executability(Exec)\end{tabular} & Development & {\ul \href{https://github.com/NVlabs/progprompt-vh}{Link}} & Yes & No & Yes & \begin{tabular}[c]{@{}l@{}} Robotic action success feedback is not\\ shared with agent leading to failure scenarios \end{tabular}   \\
61 & \citet{paper_075} & EACL & 2023 & \begin{tabular}[c]{@{}l@{}}mT5-Large,\\ PaLM\end{tabular} & Google research India & NLP & Translation & Translating English datasets to several other languages & MTOP, MASSIVE & Zero-shot and Few-shot & corrected Exact Match & Prototype & NA & NA & No & Yes & \begin{tabular}[c]{@{}l@{}} Computationally expensive\end{tabular}   \\
62 & \citet{paper_016} & ACL & 2023 & BERT, ALBERT, RoBERTa & IBM Research & Generic & Trustworthy AI & Inspect the fairness and Bias of foundation models & CrowS-Pairs & NA & NA & Conceptualization/PoC &  {\ul \href{https://github.com/IBM/finspector}{Link}} & NA & NA & Yes & \begin{tabular}[c]{@{}l@{}}The tool's effectiveness not tested for decoder-only models.\end{tabular}   \\
63 & \citet{paper_019} & EMNLP & 2023 & \begin{tabular}[c]{@{}l@{}}GPTNeo-2.7B, GPTJ-6B, \\ Open-LLaMA-7B, \\ RedPajama-7B,GPT3.5-Turbo, \\ GPT4\end{tabular} & Amazon Alexa AI & Security & Trustworthy AI & Evaluating the LLMs for Hallucinations & DBpedia, TriviaQA & Zero-shot & \begin{tabular}[c]{@{}l@{}}BLEU, ROUGE, METEOR, \\ BERTScore, AlignScore\end{tabular} & Conceptualization/PoC &  {\ul \href{https://github.com/amazon-science/invite-llm-hallucinations}{Link}} & NA & NA & NA & \begin{tabular}[c]{@{}l@{}}Lack of diversity in test set.\end{tabular}   \\
64 & \citet{paper_027} & CI & 2023 & Not disclosed & Apple & Generic & Trustworthy AI & Identify the presence of gender bias in LLMs & Proprietary & Zero-shot & NA & Conceptualization/PoC & NA & NA & NA & Yes & Results may not reflect the real gender bias   \\
65 & \citet{paper_028} & EMNLP & 2023 & ChatGPT, Alpaca & Adobe Research & Generic & Trustworthy AI & Identify the presence of gender bias in LLMs & WikiBias-Aug & Zero-shot & WEAT & Conceptualization/PoC &  {\ul \href{https://github.com/uclanlp/biases-llm-reference-letters}{Link}} & NA & Yes & Yes & \begin{tabular}[c]{@{}l@{}}Only consider binary gender when analyzing biases\end{tabular}   \\
66 & \citet{paper_056}  & NeurIPS & 2023 & \begin{tabular}[c]{@{}l@{}}OPT-350M\\ OPT-1.3B\\ OPT- 2.7B\end{tabular} & NAVER AI Lab, Parameter Lab & Generic & Trustworthy AI & Probing for PII in a given LLM & Pile & Fewshot (In-context learning) & \begin{tabular}[c]{@{}l@{}}Sentence matching,\\ Likelihood ratio\end{tabular} & Conceptualization/PoC & NA & NA & Yes/ & Yes & \begin{tabular}[c]{@{}l@{}}Evaluation dataset uses private\\information sourced from open-source datasets\end{tabular}   \\
67 & \citet{paper_066} & \begin{tabular}[c]{@{}l@{}}EMNLP System \\ demonstrations\end{tabular} & 2023 & \begin{tabular}[c]{@{}l@{}}text-davinci-003, GPT-3.5-turbo, \\ falcon-7b-instruct, llama2-13b-chat\end{tabular} & NVIDIA & Generic & Trustworthy AI & \begin{tabular}[c]{@{}l@{}}Tool kit for adding programmable guardrails for \\ conversational LLMs\end{tabular} & \begin{tabular}[c]{@{}l@{}}Anthropic Red-Teaming and\\ Helpful datasets\end{tabular} & Fewshot (In-context learning) & Accuracy & Development & {\ul \href{https://github.com/NVIDIA/NeMo-Guardrails/}{Link}} & Yes & NA & Yes & \begin{tabular}[c]{@{}l@{}}Toolkit not suitable as standalone solution\end{tabular}   \\
68 & \citet{paper_069} & \begin{tabular}[c]{@{}l@{}}EMNLP System \\ demonstrations\end{tabular} & 2023 & Generic & {\ul H2O.ai} & Generic & Trustworthy AI & \begin{tabular}[c]{@{}l@{}}Deploy and test efficiency of wide variety of LLMs \\ on private databases and documents\end{tabular} & NA & NA & NA & Deployment &  {\ul \href{https://github.com/h2oai/h2ogpt}{Link}} & Yes & Yes & Yes & \begin{tabular}[c]{@{}l@{}}Datasets, Biases and Offensiveness, Usage, Carbon footprint, \\ Hallucinations of LLMs\end{tabular}  \\
\bottomrule
\end{tabular}}
\caption{Master table of the survey with 68 research papers.}
\label{tab:check_list}
\end{sidewaystable}

\end{document}